\documentclass[11pt,a4paper]{article}
\usepackage{jheppub}
\usepackage{amsmath}
\usepackage{verbatim}
\usepackage{amsfonts,mathrsfs,latexsym,url,wasysym,float,slashed,xfrac,tipa}
\usepackage{mathtools}
\usepackage{comment}

\newcommand{\be}{\begin{equation}}
\newcommand{\ee}{\end{equation}}

\newcommand{\ba}{\begin{eqnarray}}
\newcommand{\ea}{\end{eqnarray}}
\begin{document}
\title{Black hole-de Sitter model, a proposal for the de Sitter phases}

\author{Ida M. Rasulian}
\affiliation{ School of Physics, Institute for Research in Fundamental
Sciences (IPM), P.O.Box 19395-5531, Tehran, IRAN.}
\abstract{
    We expand on the braneworld-black hole de Sitter model introduced in \cite{Rasulian} which is a proposal for constructing an effective de Sitter spacetime with no explicit dependence on brane tension or bulk cosmological constant. In this model the 4D de Sitter space emerges on a brane near the horizon of a 5D black hole. We study the effective gravity on the brane non-perturbatively, with an approximate Z2 symmetry assumption, up to a conformal factor, and find that the evolution of the effective cosmological constant on the brane depends on the flux of energy towards and away from the black hole in the bulk. In this setup the presence of the black hole horizon sets the initial condition for the brane's evolution and the brane approaches its null configuration with de Sitter length $l\lesssim l_5$, where $l_5$ is the 5D Planck's length, as soon as the horizon forms. During the last stages of collapse following this phase (or further flux of matter in the bulk after the horizon is formed), the effective de Sitter length on the brane increases due to the in-falling flux relatively fast. This phase is tentatively the transition between a low scale inflationary phase and the late dark energy phase. Also we observe that the increase in the de Sitter length is accompanied with a flux of energy entering the brane due to the jump in the bulk flux across the brane. Considering the initial state of the brane to be in the $l_5$ neighborhood of the horizon, the configuration which is slightly below the horizon is Euclidean AdS with AdS radius $l\lesssim l_5$. This can be interpreted as a boundary proposal for the resulting cosmology.}
\maketitle
\section{Introduction}
De Sitter space construction is an important problem that has been a subject of active debate in the context of string theory, as a candidate theory of quantum gravity, and highlighted in the Swampland program (see for instance \cite{Obied}\cite{Krishnan}\cite{Bedroya}). In this work we follow a different path. We propose a setup that is mainly classical in a braneworld scenario, and close to a black hole horizon, where the resulting de Sitter length does not explicitly depend on brane tension or bulk cosmological constant. 

Braneworld cosmological models have a long history and the literature in this context is vast (see for instance \cite{Battye}\cite{Sasaki}\cite{Swingle}\cite{Ida}). To the best of the author's knowledge, the effective cosmological constant that appear in these frameworks, depends explicitly on the brane tension and/or the bulk cosmological constant.

In the current work the initial condition for the value of cosmological constant is set by the presence of the black hole after the horizon is formed in the collapse process and its further evolution is a result of the flux towards the black hole during the last stages of collapse. After the collapse is settled the cosmological constant remains classically constant. During this phase, Hawking radiation can tentatively lead to lowering the de Sitter length.

In section 2 we introduce the construction. In section 3 we study gravity on the brane non-perturbatively, using an approximate Z2 symmetry assumption (Z2 symmetry up to a conformal factor). In section 4 we elaborate more on the resulting cosmological constant and its evolution which is the main goal of the current work. We conclude in section 5.
\section{The setup}
We briefly review the construction in \cite{Rasulian}.
We begin by considering the 5D Schwarzschild metric
\be\label{metric}
{\rm d}s^2=-\Big(1-\frac{l_s^2}{R^2}\Big){\rm d} t^2+\frac{{\rm d}R^2}{\Big(1-\frac{l_s^2}{R^2}\Big)}+R^2 {\rm d}\theta^2 +R^2 \sin ^2 \theta {\rm d}\Omega_2^2,
\ee
where $l_s$ is the Schwarzschild radius.

Assuming we have a brane with location parameterized by $R=f(\theta)$, the induced metric on the brane, ignoring back-reaction, will be 
\be 
{\rm d}s_{ind}^2=-\big(1-\frac{l_s^2}{f^2}\big)dt^2+\big(1-\frac{l_s^2}{f^2}\big)^{-1}f'^2d\theta^2+f^2 d\theta^2+f^2 \sin^2\theta d\Omega_2^2.
\ee 
The brane action is (in the absence of matter on the brane)
\be 
S=-\sigma \int d^4 x \sqrt{-g}.
\ee 
Substituting the induced metric in this action we find
\be
L=-\sigma(f \sin\theta)^2\sqrt{\Big | f^2-l_s^2+f'^2\Big |}.
\ee
A solution that extremizes this action, which corresponds to a null brane, is
\be 
f(\theta)=l_s \cos \theta.
\ee 
We expect that in the presence of extra stress tensor, the position of the brane will be distorted with respect to this null configuration.

From another point of view, this null configuration is the asymptotic state of the brane during the collapse process as viewed by an accelerated observer, as soon as a horizon is formed. From the point of view of the observer on the brane, we will see that below this null configuration, the geometry of the brane is Euclidean AdS and the observer on the brane that has no access to the bulk coordinate $l$, which now plays the role of time, loses the commonsense perception of time. The EAdS phase of the brane which has a varying EAdS radius plays the role of a boundary proposal for the universe in this setup. We will elaborate more on this in the next sections.
\subsection{Metric close to the null configuration}
We use the following change of coordinates (see figure 1) in \eqref{metric},
\be 
R=(l_s+\epsilon)\cos\theta.
\ee
\begin{figure}
\begin{center}
\includegraphics[scale=.5]{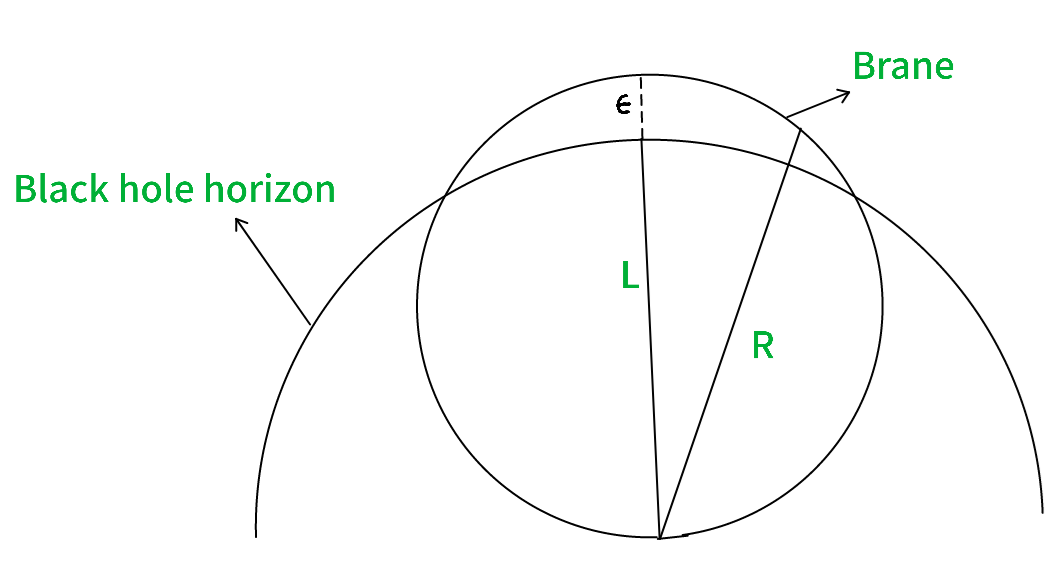}
\end{center}
\caption{The Setup Illustration}
\end{figure}
Then the metric for $\epsilon>0$ and small $\theta$ and $\epsilon/l_s$ can be brought to the form
\be\label{out1}
{\rm d}s^2=-\frac{l^2}{l_s^2}(1-\frac{r^2}{l^2})dt^2+\frac{1}{1-\frac{r^2}{l^2}}(dr^2+dl^2-2\frac{r}{l}dr dl)+r^2 d\Omega_2^2
\ee
where we have defined $r=l_s \theta$ and $l=\sqrt{2l_s \epsilon}$. In order to have a diagonal metric we further use the redefinition $r=\frac{l}{l_s}\rho$ such that the metric becomes
\be\label{in1} 
{\rm d}s^2=\frac{2\epsilon}{l_s}\Big(-(1-\frac{\rho^2}{l_s^2})dt^2+\frac{d\rho^2}{1-\frac{\rho^2}{l_s^2}}+\rho^2 d\Omega_2^2\Big)+\frac{l_s}{2\epsilon}d\epsilon^2,
\ee 
Following a similar calculation for $\epsilon<0$ we find
\be 
{\rm d}s^2=\frac{-2\epsilon}{l_s}\Big((1+\frac{\rho^2}{l_s^2})dt^2+\frac{d\rho^2}{1+\frac{\rho^2}{l_s^2}}+r^2 d\Omega_2^2\Big)+\frac{l_s}{2\epsilon}d\epsilon^2,
\ee
where we have used $r=\sqrt{\frac{l_s}{2|\epsilon|}}l_s \theta$ to derive a diagonal metric.

We see that for fixed $\epsilon<0$ the metric on a brane with constant $\epsilon$ is Euclidean AdS and for fixed $\epsilon>0$ the metric on the brane is de Sitter. The EAdS part is an interesting feature that can be considered as a boundary proposal. We can see that if the universe begins from $\epsilon\approx 0$ the boundary proposal in this setup will be a Euclidean AdS metric. 

\section{Effective gravity on the brane, a non-perturbative treatment}
The metric \eqref{in1} can be brought to the form
\ba\label{background-metric}
{\rm d}s^2&=&\frac{l^2}{l_s^2}\Big(-(1-\frac{r^2}{l_s^2})dt'^2+\frac{dr^2}{1-\frac{r^2}{l_s^2}}+r^2 d\Omega_2^2\Big)+dl^2, 
\ea

Using a change of coordinate $\frac{l}{l_s}=e^{\frac{z}{l_s}}$ this can be written as
\be 
{\rm d} s^2=e^{\frac{2z}{l_s}}(\tilde{g}_{\mu\nu}dx^\mu dx^\nu +dz^2).
\ee

In order to consider a more general case we begin with the metric ansatz
\be 
{\rm d} s^2=f^2(x^\alpha ,z)(\tilde{g}_{\mu\nu}dx^\mu dx^\nu +dz^2),
\ee
where we make $f$ general to keep the position of the brane at some fixed $z=z_0$. We assume the Z2 breaking effect lies in $f(x^\alpha , z)$ and the rest of the metric is Z2 symmetric. In particular\footnote{We use the common notation that $\langle A\rangle=\frac{A^+ +A^-}{2}$ is the mean value of $A$ across the brane and $[A]=A^+ -A^-$ is the jump in $A$ across the brane.}
\be 
\langle \partial_z \tilde{g}_{\mu\nu}\rangle=0.
\ee
Using a conformal transformation $g_{M N}=f^2 \tilde{g}_{M N}$ we move to a Z2 symmetric system such that the stress tensor in the two systems are in general related as \cite{Blaschke}
\be 
\kappa_5 T_{M N}=\kappa_5 \tilde{T}_{M N}+\frac{6}{f^2}\partial_M f \partial_N f-\frac{3}{f}\tilde{\nabla}^{(5)}_M \partial_N f+\frac{3}{f}\tilde{g}_{M N}\tilde{\square}^{(5)}f,
\ee
or specifically
\be\label{must}
\kappa_5 T_{\mu \nu}=\kappa_5 \tilde{T}_{\mu\nu}+\frac{6}{f^2}\partial_\mu f \partial_\nu f-\frac{3}{f}\tilde{\nabla}^{(4)}_\mu \partial_\nu f+\frac{3}{f}\tilde{\Gamma}^5_{\mu\nu}\partial_5 f+\frac{3}{f}\tilde{g}_{\mu\nu}(\partial^2_5 f+\tilde{\square}^{(4)} f-\tilde{\Gamma}^5 \partial_5 f),
\ee
\be\label{55st}
\kappa_5 T_{55}=\kappa_5 \tilde{T}_{55}+\frac{6}{f^2}(\partial_5 f)^2+\frac{3}{f}(\tilde{\square}^{(4)}f-\tilde{\Gamma}^5 \partial_5 f),
\ee
\be\label{mu5st} 
\kappa_5 T_{\mu 5}=\kappa_5 \tilde{T}_{\mu 5}+\frac{6}{f^2}\partial_5 f \partial_\mu f-\frac{3}{f}\partial_\mu \partial_5 f-\frac{3}{f}\tilde{\Gamma}^{5\alpha}_\mu \partial_\alpha f.
\ee

We have from the junction conditions \cite{Israel}
\be 
[\tilde{\Gamma}^5_{\mu\nu}]=-\frac{1}{2}\partial_5 \tilde{g}_{\mu\nu}=\kappa_5 (\tilde{S}_{\mu\nu}-\frac{\tilde{S}}{3}\tilde{g}_{\mu\nu}),
\ee
where $\tilde{S}_{\mu\nu}$ is the stress tensor on the brane in Z2 symmetric frame.

We can find from \eqref{must} and \eqref{55st} that
\be\label{jumpmu}
[T_{\mu\nu}]=\frac{3}{f}\partial_5 f \tilde{S}_{\mu\nu},
\ee
and 
\be 
[T_{55}]=\frac{\partial_5 f}{f}\tilde{S}.
\ee

The Einstein tensor in Z2 symmetric frame can be written as
\be\label{4d}
\kappa_5 \tilde{T}_{\mu\nu}=\tilde{G}_{\mu\nu}=\tilde{G}^{(4)}_{\mu\nu}+2\tilde{\Gamma}^{5\alpha}_\nu \tilde{\Gamma}^5_{\mu\alpha}-\tilde{\Gamma}^5\tilde{\Gamma}^5_{\mu\nu}+\partial_5 \tilde{\Gamma}^5_{\mu\nu}-\tilde{g}_{\mu\nu}\partial_5 \Gamma^5 +\frac{1}{2}\tilde{g}_{\mu\nu}(\tilde{\Gamma}^5)^2+\frac{1}{2}\tilde{g}_{\mu\nu} \tilde{\Gamma}^{5\alpha\beta}\tilde{\Gamma}^5_{\alpha\beta},
\ee
\be\label{55}
\kappa_5 \tilde{T}_{55}=-\frac{1}{2}\tilde{R}^{(4)}+\frac{1}{2}(\tilde{\Gamma}^5)^2-\frac{1}{2}\tilde{\Gamma}^{5\alpha\beta}\tilde{\Gamma}^5_{\alpha \beta}.
\ee
Taking a trace over \eqref{4d} and using \eqref{55} we find
\be\label{trace}
\kappa_5 (2\tilde{T}_{55}-\tilde{T})=-3\tilde{\Gamma}^{5\alpha \beta}\tilde{\Gamma}_{5\alpha\beta}+3\partial_5 \tilde{\Gamma}^5,
\ee
where $\tilde{T}$ is the trace over $4d$ part of stress tensor.

We need to impose a constraint on $\langle \partial_z T_{\mu\nu}\rangle$ and $\langle \partial_z T_{55}\rangle$. But for now we keep them unfixed and continue the calculation. Taking a derivative with respect to $z$ from \eqref{must} and considering the mean value we have
\begin{multline}
\kappa_5 \langle \partial_z T_{\mu\nu}\rangle = 6\frac{\partial_z f}{f}\frac{\partial_\mu f \partial_\nu f}{f^2}-3\frac{\partial_z f}{f}\frac{\nabla^{(4)}_\mu \partial_\nu f}{f}+3\tilde{g}_{\mu\nu}\frac{\partial_z f}{f}(\frac{\partial^3_z f}{\partial_z f}-\frac{\partial^2_z f}{f}+\frac{\tilde{\square}^{(4)}f}{f}+2\frac{\partial^{\tilde{\alpha}}f \partial_\alpha f}{f^2})\\ -\frac{\kappa_5^2 \partial_z f}{2f} \tilde{S}(\tilde{S}_{\mu\nu}-\frac{\tilde{S}}{3}\tilde{g}_{\mu\nu})+\frac{3\partial_z f}{f}(\langle \partial_z \tilde{\Gamma}^5_{\mu\nu}\rangle-\tilde{g}_{\mu\nu}\langle \partial_z \tilde{\Gamma}^5 \rangle)+\frac{\kappa_5}{2}(\tilde{\nabla}^{(4)}_\nu \langle T_{\mu 5}\rangle+\tilde{\nabla}^{(4)}_\mu \langle T_{\nu 5}\rangle)\\-\frac{\kappa_5}{2}(\frac{\partial_\mu f}{f}\langle T_{\nu 5\rangle}+\frac{\partial_\nu f}{f}\langle T_{\mu 5}\rangle)-\kappa_5 \tilde{g}_{\mu\nu}(\tilde{g}^{\alpha\beta}\tilde{\nabla}^{(4)}_\alpha \langle T_{\beta 5}\rangle+3\tilde{g}^{\alpha\beta}\frac{\partial_\alpha f}{f}\langle T_{\beta 5}\rangle).
\end{multline}
Taking a trace we can find
\begin{multline}\label{gamma5}
\langle \partial_z \tilde{\Gamma}^5 \rangle=-\frac{\kappa_5}{9}\frac{f}{\partial_z f}\tilde{g}^{\alpha \beta}\langle \partial_z T_{\alpha \beta}\rangle+\frac{10}{3}\frac{\partial^{\tilde{\alpha}}f \partial_\alpha f}{f^2}+\frac{\tilde{\square}^{(4)}f}{f}+\frac{4}{3}\frac{\partial^3_z f}{\partial_z f}-\frac{4}{3}\frac{\partial^2_z f}{f}-\frac{\kappa_5^2}{6}\tilde{S}^{\alpha\beta}\tilde{S}_{\alpha \beta}+\frac{\kappa_5^2}{18}\tilde{S}^2\\-\frac{\kappa_5}{3} \frac{f}{\partial_z f}\tilde{g}^{\alpha\beta}\tilde{\nabla}^{(4)}_\alpha \langle T_{\beta 5}\rangle-\frac{13}{9}\kappa_5 \frac{f}{\partial_z f}\tilde{g}^{\alpha\beta}\frac{\partial_\alpha f}{f}\langle T_{\beta 5}\rangle,
\end{multline}
and 
\begin{multline}
\langle \partial_z \tilde{\Gamma}^5_{\mu\nu}\rangle =-\frac{\kappa_5}{9}\frac{f}{\partial_z f}\tilde{g}_{\mu\nu}\tilde{g}^{\alpha \beta}\langle \partial_z T_{\alpha\beta}\rangle+\frac{\kappa_5}{3}\frac{f}{\partial_z f}\langle \partial_z T_{\mu\nu}\rangle+\frac{4}{3}\frac{\partial^{\tilde{\alpha}}f \partial_\alpha f}{f^2}\tilde{g}_{\mu\nu}+\frac{1}{3}\frac{\partial^3_z f}{\partial_z f}\tilde{g}_{\mu\nu}-\frac{1}{3}\frac{\partial^2_z f}{f}\tilde{g}_{\mu\nu}\\-2\frac{\partial_\mu f \partial_\nu f}{f^2}+\frac{\tilde{\nabla}^{(4)}_\mu\partial_\nu f}{f}+\frac{\kappa_5^2}{6}\tilde{S}\tilde{S}_{\mu\nu}-\frac{\kappa_5^2}{6} \tilde{S}^{\alpha\beta}\tilde{S}_{\alpha\beta}\tilde{g}_{\mu\nu}-\frac{\kappa_5}{6}\frac{f}{\partial_z f}(\tilde{\nabla}^{(4)}_\nu \langle T_{\mu 5}\rangle+\tilde{\nabla}^{(4)}_\mu \langle T_{\nu 5}\rangle)\\+\frac{\kappa_5}{6}\frac{f}{\partial_z f}(\frac{\partial_\mu f}{f}\langle T_{\nu 5}\rangle+\frac{\partial_\nu f}{f}\langle T_{\mu 5}\rangle)-\frac{4\kappa_5}{9}\frac{f}{\partial_z f}\tilde{g}^{\alpha\beta}\frac{\partial_\alpha f}{f}\langle T_{\beta 5}\rangle\tilde{g}_{\mu\nu}.
\end{multline}
Taking a derivative with respect to $z$ in \eqref{55st} and using \eqref{gamma5} we find
\begin{multline}\label{alphaf}
\kappa_5 \langle \partial_z T_{55}\rangle=16 \frac{\partial_z f}{f}\frac{\partial^2_5 f}{f}-4\frac{\partial^3_z f}{f}-12(\frac{\partial_z f}{f})^3-4\frac{\partial_z f}{f}\frac{\partial^{\tilde{\alpha}}f \partial_\alpha f}{f^2}+\frac{\kappa_5}{3}\tilde{g}^{\alpha \beta}\langle \partial_z T_{\alpha \beta}\rangle \\+\frac{\kappa_5^2}{2}\frac{\partial_z f}{f}\tilde{S}^{\alpha \beta}\tilde{S}_{\alpha \beta}-\frac{\kappa_5^2}{6}\frac{\partial_z f}{f}\tilde{S}^2-\frac{4\kappa_5}{3}\tilde{g}^{\alpha\beta}\frac{\partial_\alpha f}{f}\langle T_{\beta 5}\rangle.
\end{multline}
We can also relate the original 4d Einstein tensor to the Z2 symmetric 4d Einstein tensor as
\be 
G^{(4)}_{\mu\nu}=\tilde{G}^{(4)}_{\mu\nu}+\frac{1}{f^2}(4\partial_\mu f\partial_\nu f-\tilde{g}_{\mu\nu}\partial^{\tilde{\alpha}}f \partial_\alpha f)-\frac{2}{f}(\tilde{\nabla}^{(4)}_\mu\partial_\nu f-\tilde{g}_{\mu\nu}\tilde{\square}^{(4)}f).
\ee
From \eqref{4d}, \eqref{must} and \eqref{alphaf} we can find considering the mean value
\begin{multline}\label{effect}
G^{(4)}_{\mu\nu}=\kappa_5 \langle T_{\mu\nu}\rangle-\frac{\kappa_5}{4}\frac{f}{\partial_z f}\langle \partial_z T_{55}\rangle \tilde{g}_{\mu\nu}+\frac{\kappa_5}{12}\frac{f}{\partial_z f}\tilde{g}^{\alpha \beta}\langle \partial_z T_{\alpha \beta}\rangle \tilde{g}_{\mu\nu}-\frac{\kappa_5}{3}\frac{f}{\partial_z f}\langle \partial_z T_{\mu\nu}\rangle \\+\frac{\kappa_5}{6}\frac{f}{\partial_z f}(\tilde{\nabla}^{(4)}_\nu \langle T_{\mu 5}\rangle+\tilde{\nabla}^{(4)}_\mu \langle T_{\nu 5}\rangle)-\frac{\kappa_5}{6}\frac{f}{\partial_z f}(\frac{\partial_\mu f}{f}\langle T_{\nu 5}\rangle+\frac{\partial_\nu f}{f}\langle T_{\mu 5}\rangle)-\frac{4\kappa_5}{3}\frac{f}{\partial_z f}\tilde{g}^{\alpha\beta}\frac{\partial_\alpha f}{f}\langle T_{\beta 5}\rangle\\-\frac{\kappa_5}{3}\frac{f}{\partial_z f}\tilde{g}^{\alpha \beta}\tilde{\nabla}^{(4)}_\alpha 
\langle T_{\beta 5}\rangle\tilde{g}_{\mu\nu}-3(\frac{\partial_z f}{f^2})^2g_{\mu \nu}-\frac{\kappa_5^2}{2}\tilde{S}^\alpha_\nu \tilde{S}_{\mu \alpha}+\frac{\kappa_5^2}{12}\tilde{S}\tilde{S}_{\mu\nu}.
\end{multline}

We observe that effective gravity on the brane explicitly depends on the behaviour of stress tensor and its first order derivative with respect to $z$, in the bulk.\footnote{
We also have from the 5'th component of 5 stress tensor conservation
\be
\tilde{\nabla}^{(4)\alpha} \langle T_{\alpha 5}\rangle+3\frac{\tilde{\partial}^\alpha f}{f}\langle T_{\alpha 5}\rangle+\partial_z T_{55}+\tilde{\Gamma}^{5\alpha \beta}T_{\alpha\beta}+\frac{\partial_z f}{f}(2T_{55}-\hat{T})-\tilde{\Gamma}^5 T_{55}=0,
\ee
where $\hat{T}=f^2 T$.} For this reason here we do not attempt to impose more constraints and assumptions on the bulk stress tensor, leaving the tentative form of the bulk stress tensor which can have a quantum or classical origin to a future work.
\section{Effective cosmological constant}
From \eqref{effect} we observe that we have an effective cosmological constant term on the brane given by $-3\Big(\frac{\partial_z f}{f^2}\Big)^2g_{\mu\nu}$. From \eqref{mu5st} and taking the mean value on both sides, we can write
\begin{equation}\label{evolv}
    3\partial_\mu (\frac{\partial_z f}{f^2})=-\frac{\kappa_5}{f} \langle T_{\mu 5}\rangle.
\end{equation}
Therefore our cosmological constant term will be constant if we assume $\langle T_{\mu 5}\rangle=0$. Otherwise we can have a decreasing or increasing cosmological constant if we have a flux respectively towards or away from the black hole.

We note that the initial value for the evolution of the cosmological constant term can be fixed considering the discussion in section 2, where we impose that in the absence of stress tensor in the bulk and on the brane (hence no flux), the brane is stabilized with a very small de Sitter length that corresponds to a very large cosmological constant. We can put a bound on how small the de Sitter length can be by approximating it with the 5D Planck's length.

From \eqref{effect} and \eqref{jumpmu} we can separate the bulk stress tensor $T_{\mu\nu}$ as
\begin{equation}\label{darkm}
    \langle T_{\mu\nu}\rangle=T^-_{\mu\nu}+\frac{3}{2}\frac{\partial_z f}{f^2}S_{\mu\nu},
\end{equation}
which is motivated by requiring both contributions to satisfy the conventional energy conditions. 
From this we can write
\begin{equation}
    \kappa_4\propto \frac{\partial_z f}{f^2}\kappa_5,
\end{equation}
where $\kappa_4$ is the effective gravitational constant on the brane. Given the current value for the 4D cosmological constant and the current de Sitter length we can find 
\begin{equation}
    l_5\approx 10^{20}l_4,
\end{equation}
where $l_5$ and $l_4$ are 5D and 4D Planck lengths.

Therefore the 5D Planck's mass is $m_5\approx 100 MeV$. The effective UV cut-off in the bulk corresponds to an energy density of order $m_5^5$ which due to \eqref{jumpmu} leads to an approximated energy density on the brane of order $l_0 m_5^5=m_{UV}^4$, which results $m_{UV}\approx 10^6 TeV.$\cite{Rasulian}
\subsection{The inflationary phase}
Considering the redefinition
\be 
R=L \cos\theta
\ee
the exact metric (in the absence of stress tensor) for $L>l_s$ becomes
\begin{multline}\label{out}
{\rm d}s_{out}^2=\frac{-l_{out}^2}{l_s^2+l_{out}^2}(1-\frac{r^2}{l_s^2})dt^2+\frac{l_{out}^2(l_s^2+l_{out}^2)}{l_s^4}\frac{dr^2}{(1-\frac{r^2}{l_s^2})(1+\frac{r^2 l_{out}^2}{l_s^4})^2}\\+\frac{(l_s^2+l_{out}^2)r^2 l_{out}^2}{l_s^4}\frac{d\Omega_2^2}{(1+\frac{r^2 l_{out}^2}{l_s^4})^2}+dl_{out}^2,
\end{multline}
where we have defined $l_{out}^2=L^2-l_s^2$. For the case of $L<l_s$ we have
\begin{multline}
{\rm d}s_{in}^2=\frac{l_{in}^2}{l_s^2-l_{in}^2}(1+\frac{r^2}{l_s^2})dt^2+\frac{l_{in}^2(l_s^2-l_{in}^2)}{l_s^4}\frac{dr^2}{(1+\frac{r^2}{l_s^2})(1+\frac{r^2 l_{in}^2}{l_s^4})^2}\\+\frac{(l_s^2-l_{in}^2)r^2 l_{in}^2}{l_s^4}\frac{d\Omega_2^2}{(1+\frac{r^2 l_{in}^2}{l_s^4})^2}-dl_{in}^2,
\end{multline}
where $l_{in}^2=l_s^2-L^2$, which for $l_{out/in}\ll l_s$ tend to \eqref{out1} and \eqref{in1}.

We can see that in the second case $l_{in}$ plays the role of time inside the black hole and the metric on the constant $l_{in}$ brane, which is a constant time slice inside the black hole, is Euclidean AdS with a $l_{in}$ varying AdS length for $l_{in}\ll l_s$ and the Ricci scalar becomes infinite on the brane as $l_{in}$ tends to $l_s$.

Consider the approximately null configuration with $l\lesssim l_5$.
This corresponds to a very large cosmological constant. According to \eqref{evolv} if there is a flux of energy towards the black hole, the brane in this configuration evolves (in the time parameter related to the outside of black hole) such that the cosmological constant is reduced. (See figure 2.)
\begin{figure}
    \centering
    \includegraphics[width=0.5\linewidth]{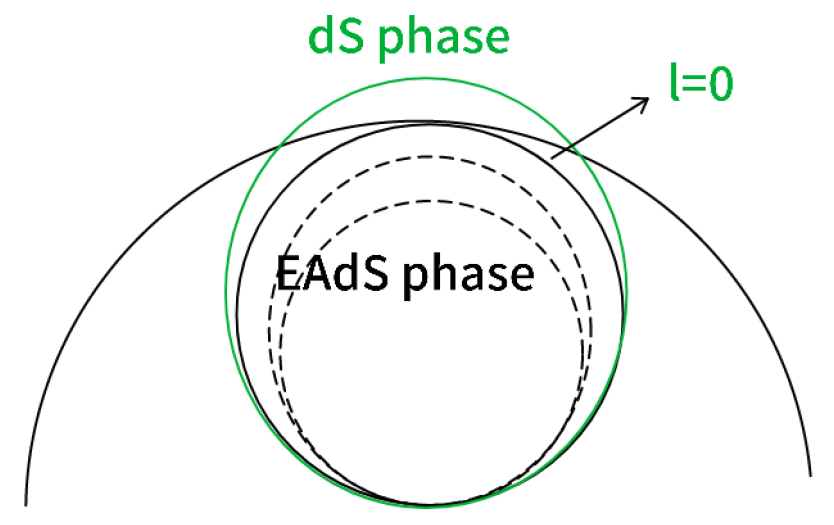}
    \caption{The universe emerges from the black hole at $l\approx 0$. The EAdS phase can be considered as a boundary.}
    \label{fig:enter-label}
\end{figure}

We find that given a nonzero value for the mean value of the energy flux $\langle T^{\mu 5}\rangle$ we necessarily have a nonzero jump $[T^{\mu 5}]$ across the brane that violates energy conservation on the brane. In particular we observe that when the flux is towards the black hole (which decreases the effective cosmological constant term),  the energy flows to the brane and the amount of matter on the brane is effectively increased.

We begin from bulk stress tensor conservation $\nabla_M T^{MN}=0$. Setting $N=\mu$ one of the brane coordinates, considering the delta-function contribution, we have
\begin{equation}\label{branecons}
   \nabla^{(4)}_\alpha S^{\alpha \mu}=-[T^{5\mu}].
\end{equation}
Next we want to study the jump in this equation. We have from \eqref{mu5st},
\begin{multline}
    \kappa_5 [\partial_5 T^{5\mu}]=-4\frac{\partial_z f}{f}\kappa_5 [T^{5\mu}]+(-9\frac{\partial_\alpha\partial_z f}{f^5}+15 \frac{\partial_\alpha f\partial_5 f}{f^6})\kappa_5(\tilde{S}^{\mu\alpha}-\frac{\tilde{S}}{3}\tilde{g}^{\mu\alpha})\\=-4\kappa_5 \frac{\partial_z f}{f}[T^{\mu 5}]-\frac{3\partial_z f\partial_\alpha f}{f^6}\kappa_5 (\tilde{S}^{\mu\alpha}-\frac{\tilde{S}}{3}\tilde{g}^{\mu\alpha})+\frac{3\kappa_5^2}{f^4}\langle T_{\alpha 5}\rangle (\tilde{S}^{\mu \alpha}-\frac{\tilde{S}}{3}\tilde{g}^{\mu\alpha}),
\end{multline}
where in the second line we used \eqref{evolv}. Using the conservation equation again and considering the jump we find after some calculation
\begin{equation}
    12\frac{\partial_z f}{f^2}[T^{\mu 5}]+\kappa_5 S\langle T^{\mu 5}\rangle=0.
\end{equation}

For instance during the last stages of collapse where $\langle T^{0 5}\rangle<0$, we have $[T^{0 5}]<0$ and according to \eqref{branecons} we have $\nabla^{(4)}_\alpha S^{\alpha 0}>0$ which means there is a flux of energy to the brane and this increases the amount of energy on the brane during our tentative inflationary period.
\section{Discussion}
We proposed a cosmological brane-world scenario, as a potential model that can explain the de Sitter phases of our universe. In this setup the initial condition of the universe is set by the presence of the black hole horizon, and evolves from a EAdS past to a de Sitter configuration with $l\lesssim l_5$ and later to the dark energy phase with large de Sitter length.

We envisage that during the collapse process which forms this bulk black hole, as soon as the horizon forms the brane enters its inflationary phase with $l_{dS}\lesssim l_5\approx 10^{20}l_P$, where $l_5$ and $l_P$ are respectively the 5D and 4D Planck's lengths. As the collapse continues to its last stages the flux towards the black hole decreases the effective cosmological constant on the brane, meanwhile leading to a flow of energy to the brane. After the collapse process is finalized the cosmological constant will approximately remain constant. We expect that effects like Hawking radiation can slowly increase the effective cosmological constant on the brane in late time.

In the future we will study the bulk stress tensor from classical and quantum points of view to find the explicit form of gravity on the brane. The role of $T^-_{\mu\nu}$ (which is the bulk stress tensor below the brane) in \eqref{darkm} which contributes linearly to 4D Einstein equations needs to be understood in more detail, since it has the potential of being related to dark matter.
\section*{Acknowledgement}
I thank Amjad Ashoorioon for comments and discussions.

\end{document}